\documentclass[conference]{IEEEtran}
\IEEEoverridecommandlockouts
\usepackage{cite}
\usepackage{amsmath,amssymb,amsfonts}
\usepackage{algorithmic}
\usepackage{graphicx}
\usepackage{textcomp}
\usepackage{hyperref}
\usepackage{subcaption}
\usepackage{xcolor}
\def\BibTeX{{\rm B\kern-.05em{\sc i\kern-.025em b}\kern-.08em
    T\kern-.1667em\lower.7ex\hbox{E}\kern-.125emX}}
\begin{document}

\title{Topological Analysis of Mixer Activities in the Bitcoin Network}

\author{\IEEEauthorblockN{Francesco Zola}
\IEEEauthorblockA{\textit{Digital Security,} \\
\textit{Vicomtech Foundation}\\
Donostia, Spain \\
fzola@vicomtech.org}
\and
\IEEEauthorblockN{Jon Ander Medina}
\IEEEauthorblockA{\textit{Digital Security,}\\
\textit{Vicomtech Foundation}\\
Donostia, Spain\\
jamedina@vicomtech.org}
\and
\IEEEauthorblockN{Andrea Venturi}
\IEEEauthorblockA{\textit{Digital Security,}\\
\textit{Vicomtech Foundation}\\
Donostia, Spain\\
aventuri@vicomtech.org}
\and
\IEEEauthorblockN{Raul Orduna}
\IEEEauthorblockA{\textit{Digital Security,}\\
\textit{Vicomtech Foundation}\\
Donostia, Spain\\
rorduna@vicomtech.org}}

\maketitle

\begin{abstract}
Cryptocurrency users increasingly rely on obfuscation techniques such as mixers, swappers, and decentralised or no-KYC exchanges to protect their anonymity. However, at the same time, these services are exploited by criminals to conceal and launder illicit funds. Among obfuscation services, mixers remain one of the most challenging entities to tackle. This is because their owners are often unwilling to cooperate with Law Enforcement Agencies, and technically, they operate as `black boxes'. 
To better understand their functionalities, this paper proposes an approach to analyse the operations of mixers by examining their address-transaction graphs and identifying topological similarities to uncover common patterns that can define the mixer's \textit{modus operandi}. The approach utilises community detection algorithms to extract dense topological structures and clustering algorithms to group similar communities. The analysis is further enriched by incorporating data from external sources related to known \textit{Exchanges}, in order to understand their role in mixer operations. The approach is applied to dissect the \textit{Blender.io} mixer activities within the Bitcoin blockchain, revealing: i) consistent structural patterns across address-transaction graphs; ii) that \textit{Exchanges} play a key role, following a well-established pattern, which raises several concerns about their AML/KYC policies. This paper represents an initial step toward dissecting and understanding the complex nature of mixer operations in cryptocurrency networks and extracting its \textit{modus operandi}.
\end{abstract}


\section{Introduction}

Today, we are witnessing an increasing emergence of High-Risk Criminal Networks, where especially organised groups, but also terrorists, engage in criminal activities that can rapidly reach a large population and generate a significant impact on public safety, economic stability, or national security \cite{cepol1}. At the same time, emerging communication platforms and various tactics including the Crime-as-a-Service (CaaS) strategy, allow inexperienced users to access and use dedicated criminal services such as ransomware creator, anonymisation networks, phishing kits, money mule \cite{europol2024iocta}. In both cases, whether involving organised groups or novice cybercriminals, the primary objective remains the same: maximising the impact of attacks while concealing their tracks and laundering illicit proceeds.

In this scenario, cryptocurrency plays a pivotal role in these operations due to its anonymity, decentralisation, and borderless nature, making it a preferred medium for concealing illicit funds and facilitating money laundering. While studies \cite{jourdan2018characterizing,meiklejohn2013fistful} show that Law Enforcement Agencies (LEAs) can reduce anonymity through the 'follow-the-money' approach, criminals counteract this by employing obfuscation techniques such as money mules and international bank transfers—are often combined with crypto-specific services like mixers, swappers, and no-KYC exchanges \cite{chainalysis2024crypto,europol2024iocta}. The complexity of these methods varies based on the crime type, the cryptocurrency used, and perpetrators’ expertise \cite{europol2024iocta}.

Among dedicated services, mixers remain one of the most challenging entities to tackle. This is because their owners are often unwilling to cooperate with LEAs, and technically, they operate as `black boxes'. In fact, they leverage advanced cryptographic primitives such as hash functions (for data integrity and unlinkability), zero-knowledge proofs (to ensure anonymity), and commitment schemes (to secure and obscure transaction details) \cite{ziegeldorf2018secure}. Together, these techniques make it exceptionally difficult for investigators to trace the flow of the funds \cite{europol2023iocta}. To tackle these challenges, financial authorities such as the US Office of Foreign Assets Control (OFAC)\footnote{https://ofac.treasury.gov/}, Office of Financial Sanctions Implementation (OFSI)\footnote{https://sanctionssearchapp.ofsi.hmtreasury.gov.uk/}, European External Action Service (EEAS)\footnote{https://www.eeas.europa.eu/}, United Nations Security Council (UNSC)\footnote{https://www.un.org/securitycouncil/content/un-sc-consolidated-list}, have started imposing sanctions on mixers involved in illicit activities to discourage others from engaging relations with them. \textit{Sinbad, Tornado Cash,} and \textit{Blender.io} \cite{ofac1,ofac2,ofac3} are just a few examples of mixers sanctioned between 2022 and 2023. While this sanctioning strategy decreased funds sent to mixers from illicit addresses, it did not stop user demand.  In fact, after that, organised groups like the Lazarus Group have created their own mixer, while former users of sanctioned or shutdown mixers have migrated to new entities such as \textit{YoMix} \cite{chainalysis2024crypto}.

Recognising the challenge in understanding how funds are blended \cite{ziegeldorf2018secure,moser2017anonymous,bonneau2014mixcoin}, and the difficulty in halting mixers' activities, this work aims to take a first step toward dissecting mixer operations. Specifically, it starts mapping mixer activities within address-transaction graphs (or mixer activity graphs), and analysing topological similarities in order to extract common patterns that can be used to define the mixer's \textit{modus operandi}. This task is achieved by extracting dense topological structures in all the mixer activity graphs using community detection algorithms, and subsequently grouping similar communities using clustering algorithms. The analysis also incorporates information about other known \textit{Exchange} to understand their role in mixer activities, enriching the findings. This work aims to offer an intriguing yet partial understanding of the mixer \textit{modus operandi}. 

Information about \textit{Blender.io} mixer is used in this work to obtain a preliminary validation of the approach. \textit{Blender.io} was chosen as it was one of the first mixers to be sanctioned by the OFAC in May 2022 \cite{ofac1}. This mixer primarily operated on the Bitcoin network and was sanctioned for engaging in malicious cyber-enabled activities, including money laundering. Information about the sanctioned mixer addresses is available only from the US OFAC, as other agencies (OFSI, EEAS, etc.) either lack such information or do not release it publicly.

In summary, the main contributions of this work are: 1) introducing an initial step toward the extraction of cryptocurrency mixers' \textit{modus operandi}, 2) validating this initial step through a specific use case; 3) highlighting the pivotal role of \textit{Exchange} entities in mixer operations.

\section{Related Work}\label{sec:relatedwork}
Early researches were primarily focused on developing innovative mixing strategies. Significant examples in literature can be found in \cite{bonneau2014mixcoin,heilman2017tumblebit,ziegeldorf2018secure}, with some of them being implemented as a public service. 
On the other hand, the literature focusing on analysing and studying the behaviour of existing mixing services remains relatively sparse. A foundational work in this area is provided by Moser et al. in \cite{moser2013inquiry}, where the authors conducted the first empirical study to evaluate the extent to which Bitcoin mixers enhance user anonymity. Subsequent research has provided more detailed analyses, often with an emphasis on detecting mixing entities (e.g., transactions and addresses) in the blockchain. For example, \cite{yanovich2016shared,wu2021towards} propose heuristics aimed at identifying mixing transactions. Other research has been directed to uncover security flaws in mixing services. For example, \cite{pakki2021everything} highlights critical privacy and security issues, concluding that many mixing services fail to adopt security solutions proposed by the academic community and are often affected by severe vulnerabilities

Recent works have explored machine learning and deep learning approaches to detect mixing services \cite{xu2023find,sun2022lstm,wu2021detecting}. The intuition behind these approaches is that mixing services exhibit distinctive patterns that classifiers or clustering algorithms can identify. For instance, the work in \cite{wu2021detecting} uses statistical features from recurring topological patterns to detect mixer addresses. An interesting analysis, aimed at shedding light on mixer's \textit{modus operandi}, is presented in \cite{wu2021towards}. The authors categorise mixers into two primary groups based on their employed mechanisms: swapping and obfuscating. Swapping mechanisms utilise potentially lengthy peeling chains to exchange inputs and outputs among users, whereas obfuscating mechanisms rely on anonymity sets—outputs with identical transaction values directed to distinct users. Furthermore, the authors primarily focus on obfuscating mechanisms, presenting heuristics for their detection. Inspired by these works, this paper proposes enriching the detection of the mixer's \textit{modus operandi} by analysing topological information to highlight structural repetitions and common patterns.

\section{Graph-Based Approach}\label{sec:approach}
In Section \ref{subsec:grph}, the procedure to create the address-transaction graphs is detailed, while Section \ref{subsec:topological} introduce the methodology applied in this study.
\subsection{Address-Transaction Graph}\label{subsec:grph}
\begin{figure}
    \centering
    \includegraphics[width=\linewidth]{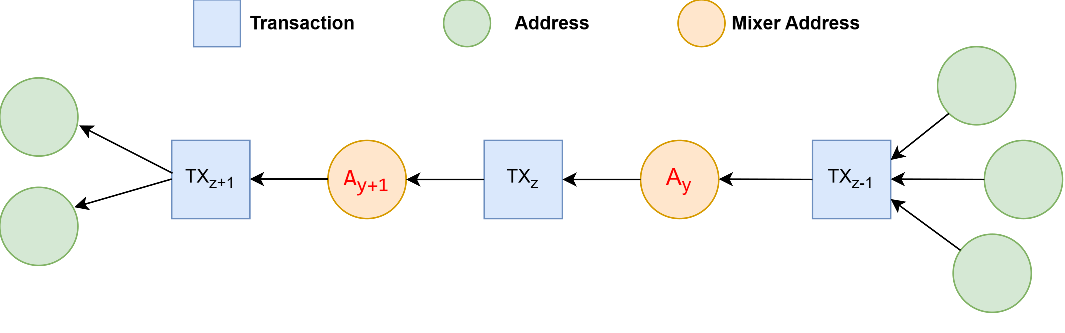}
    \caption{Example of modified 1-step address-transaction graph.}
    \label{fig:addtx}
\end{figure}

Blockchain analyses often exploit the intrinsic structure generated by transactions to build an address-transaction graph \cite{fleder2015bitcoin,jourdan2018characterizing,zola2024unveiling}. This graph is a directed graph where nodes can represent either blockchain addresses and transactions, the directed edges (arrows) from addresses to transactions represent the sender relations, and edges from transactions to addresses are receiver relations, as shown in Figure \ref{fig:addtx}. Moreover, both nodes and edges can be enriched with additional attributes such as labels, amounts, fees, timestamps, etc. 

To build this graph, a starting point needs to be defined, as well as the number of exploring steps $n$. This parameter specifies the number of transactions (both backward and forward) to explore from a selected starting point. Therefore, the graph will include all paths originating from or leading to the starting point, with a maximum length of $2n$.

In this study, \textit{Blender.io} addresses obtained from the OFAC list are used as starting points and the exploring step $n$ is set to 2. Furthermore, to better understand the relationships between mixer addresses, a modification to the graph creation process is proposed. If during the creation of the address-transaction graph of a mixer address $A\textsubscript{y}$, another mixer address $A\textsubscript{y+1}$ is discovered, $A\textsubscript{y+1}$ is used in turn as a new starting point within the same graph, as shown in Figure \ref{fig:addtx} in case of 1-step graph. In this way, we reduce the number of distinct graphs created while simultaneously retaining all the necessary information for evaluating relationships between mixer addresses within a single graph. Finally, nodes are enriched (when available) with external information, i.e., labels that identify them as belonging to known \textit{Exchanges}.

\subsection{Topological Analysis}\label{subsec:topological}
In this study, we propose conducting a topological analysis of the previously created address-transaction graph. The goal is to identify recurring patterns across multiple graphs, confirming that mixer activities follow repeated dynamics and interactions within the graph. 
For this reason, once the graphs are created (Section \ref{subsec:grph}), a community detection technique is used to split dense topological structures within all the mixer address-transaction graphs. Specifically, the Louvain Community (LC) detection algorithm is used, since it is based on maximising modularity, a measure of how dense the connections are within communities compared to those between communities \cite{blondel2024fast}. Yet, optimising this value yields the best possible grouping of nodes in a network. Several studies \cite{Singh02102020,que2015scalable} have shown that the LC algorithm consistently performs well across different networks with different complexity. 

Once the communities are extracted, 8 graph properties are extracted to describe each community topology. Four of these properties are at the graph level, providing a single value for the entire community, whereas the other four are node-level, meaning values are calculated for each node and then averages are used. The graph-level properties are: \textit{number of addresses}, \textit{number of transactions}, \textit{transitivity}, and \textit{diameter}, while the node-level properties include: \textit{degree centrality}, \textit{closeness}, \textit{betweenness}, and \textit{harmonic centrality}. These metrics are selected by combining the outcomes of previous works \cite{medina2024graphaviour,avarikioti2019ridelightninggametheory,akcora2019bitcoinheist}. Then, these topological properties are used as input to clustering algorithms to group communities that share similar properties. Specifically, two density-based clustering algorithms are used in this paper: the Ordering Points To Identify the Cluster Structure (OPTICS \cite{ankerst1999optics}) and the Hierarchical Density-Based Spatial Clustering of Applications with Noise (HDBSCAN \cite{campello2013density}). The first one works by defining dense regions using a minimum number of points that define a cluster (\textit{minPts}) and the maximum distance from one point to another for both to be considered neighbours ($\epsilon$). The second algorithm still requires the \textit{minPts} parameter but it is able to automatically tune the $\epsilon$ parameter.

Finally, once the clusters are generated, they are dissected, looking for the known entities (\textit{Exchange}) within, generalising their involvement and role in mixer operations.

\section{Data Overview \& Configuration}\label{sec:exp}
As mentioned, the main objective of this work is to dissect the behaviour of the mixer \textit{Blender.io}. At the paper date (Dec. 2024), the OFAC \textit{Specially Designated Nationals and Blocked Persons} list includes information about 45 \textit{Blender.io} addresses only related to the Bitcoin network. Since the exact names of these addresses are not relevant and do not affect the aim of this work, an identifier (ID) from 1 to 45 has been assigned to each of them and used during the analysis. Furthermore, to create the address-transaction graphs, the entire Bitcoin blockchain data is considered, i.e., more than 870k blocks and 1.120B transactions (at paper date). Finally, to enrich the address-transaction graph with known real-world entities, labelled/tagged addresses are gathered from multiple sources, such as WalletExplorer and the GraphSense tagpacks \cite{Haslhofer:2021a}. These sources have been used as "ground-truth" in many previous researches \cite{tovanich2023fingerprinting,zola2019cascading}, and allowed us to gather more than 38M addresses of almost 400 entities labelled as \textit{Exchanges, Gambling, Marketplaces, Mining Pools, Mixers, Services, Trading platforms, eWallet} and \textit{Ransomware}.

\begin{figure}[!htbp]
\centering
   \includegraphics[width=0.8\linewidth]{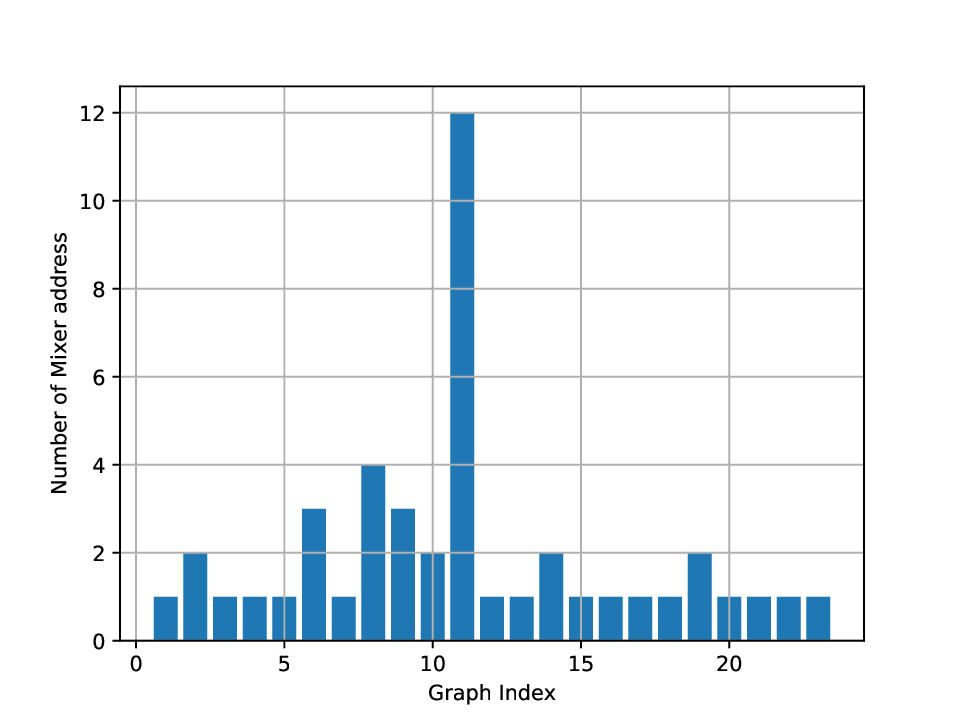}
    \caption{Number of mixer addresses in each address-transaction graph created.}
    \label{fig:hist}
\end{figure}

For the LC algorithm, the weights of the edges (e.g., the amounts sent or received) within the address-transaction graphs are considered during community generation, while other parameters like resolution and threshold are set to default values, 1 and 1e-07, respectively. Furthermore, for our analysis, all communities must be defined by address nodes at the boundaries, meaning only address nodes can have an input or output degree of 0. Therefore, we made a slight modification to the LC detection process to incorporate this requirement.

Regarding the clustering algorithms, for both the HDBSCAN and OPTICS, the minimum number of samples for creating a cluster (\textit{minPts}) is set to 5. Then, a study is conducted to evaluate how the $\epsilon$ parameter affects the OPTICS clusters, while HDBSCAN does not require such a study, as it determines the optimal value internally. Specifically, 8 different values for $\epsilon$ (0.1, 0.5, 0.9, 1, 1.5, 2, 3, 5) are tested.

\section{Preliminary Results}\label{sec:results}
Using the graph creation approach introduced in Section \ref{subsec:grph}, the 45 \textit{Blender.io} addresses generate 23 distinct address-transaction graphs, as shown in Figure \ref{fig:hist}. Yet, in one specific case, 12 mixer addresses are linked within the same graph.

\begin{figure*}[!htbp]
\centering
   \begin{subfigure}[b]{0.41\linewidth}
   \includegraphics[width=\linewidth]{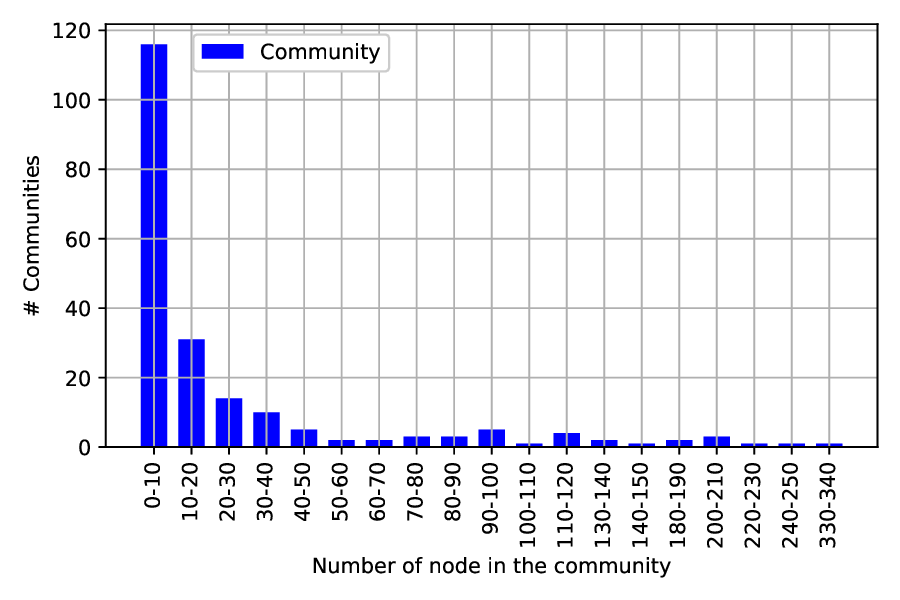}
    \caption{Histogram of number of nodes in communities.}
    \label{fig:lcresults}
\end{subfigure}
   \begin{subfigure}[b]{0.41\linewidth}
   \includegraphics[width=\linewidth]{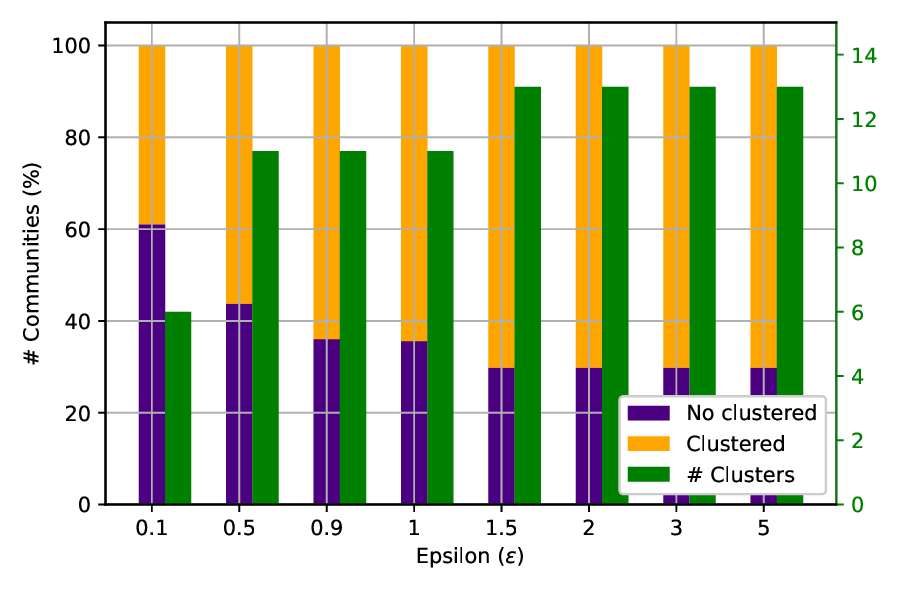}
    \caption{OPTICS clustering results with different $\epsilon$ values.}
    \label{fig:optics}
\end{subfigure}
   \begin{subfigure}[b]{0.41\linewidth}
   \includegraphics[width=\linewidth]{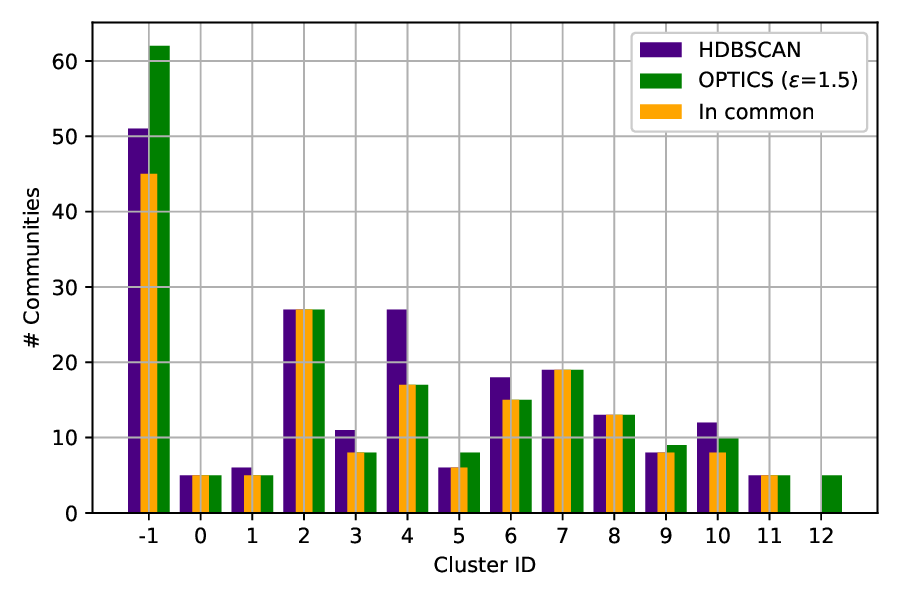}
    \caption{OPTICS vs HDBSCAN clusters (outliers ID -1).}
    \label{fig:cluster_similarities}
\end{subfigure}
    \begin{subfigure}[b]{0.41\linewidth}
   \includegraphics[width=\linewidth]{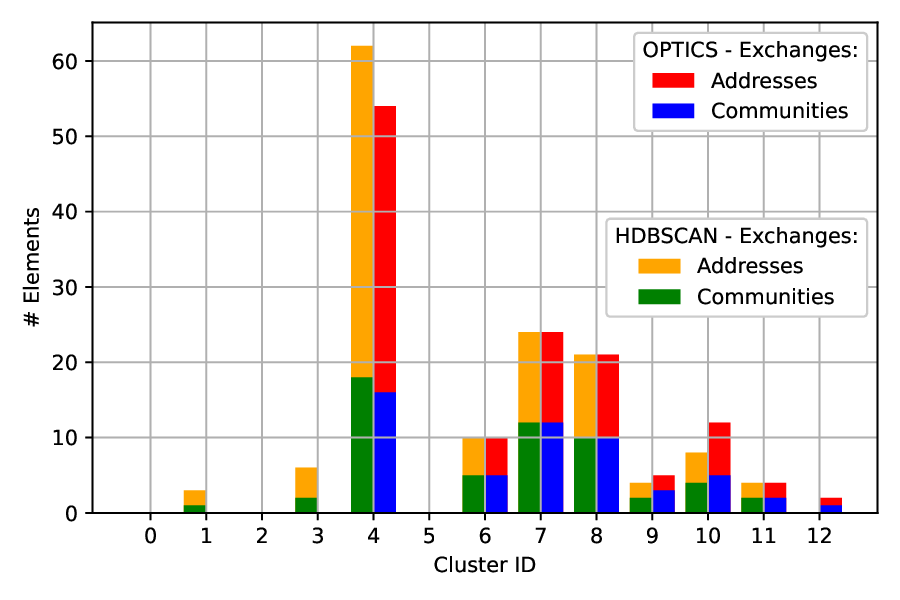}
    \caption{\textit{Exchange} entities in each cluster.}
    \label{fig:relevant_cluster_id}
  \end{subfigure}
    \caption{Preliminary results obtained using HDBSCAN and OPTICS clustering algorithms.}
    \label{fig:clustering}
\end{figure*}
\begin{figure}[!htbp]
\centering
   \includegraphics[width=0.8\linewidth]{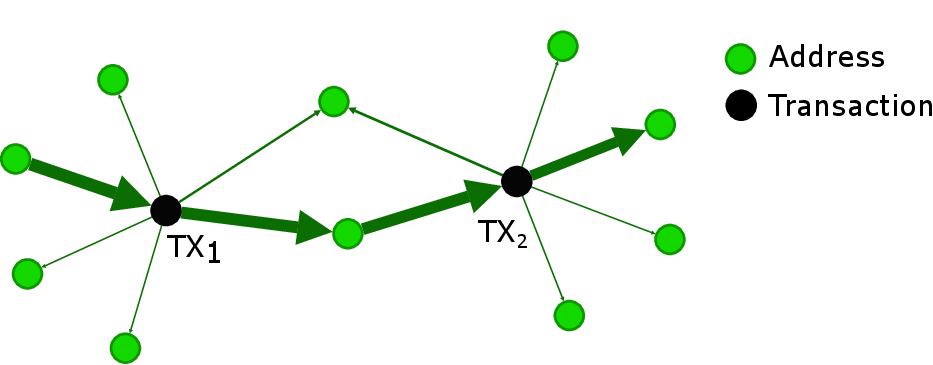}
    \caption{Predominant topologies within cluster IDs 4. The width of the edges indicates the amount sent/received.}
    \label{fig:cl4}
\end{figure}
The LC-modified algorithm (Section \ref{subsec:topological}) extracts 208 structures from the 23 address-transaction graphs. Figure \ref{fig:lcresults} shows that most of them (116) are characterised by less than 10 nodes (both address and transaction types), while just a few structures link more than 200 nodes.
Figure \ref{fig:optics} reports the performance of the OPTICS algorithm with the different $\epsilon$. The figure illustrates that as $\epsilon$ increases, more elements are assigned to clusters, leading to a decrease in the number of outliers. As a result, the number of distinct clusters grows. However, this trend holds only up to a certain threshold ($\epsilon$ = 1.5), beyond which no further changes occur in the number or composition of clusters. This threshold marks the point where the maximum number of communities is clustered, making it the reference value for comparing OPTICS with HDBSCAN clusters. Figure \ref{fig:cluster_similarities} and Figure \ref{fig:relevant_cluster_id} highlight the similarity in the clusters' creation between the two techniques. Indeed, not only the clustering algorithms are aligned in terms of elements per cluster (Figure \ref{fig:cluster_similarities}), but also in terms of labelled entities within (Figure \ref{fig:relevant_cluster_id}). The main differences lie in four aspects: i) the overall number of clusters (12 and 13 for HDBSCAN and OPTICS, respectively), ii) the number of outlier elements (51 vs 62), iii) the composition of cluster ID 4, and iv) the number of \textit{Exchange} addresses in cluster IDs 4, 9, and 10.

Another interesting finding is a large number of \textit{Exchanges} grouped in cluster IDs 4, 7, and 8, with more than 50, 12, and 10 addresses, respectively. This highlights not only their involvement in mixer activities but also their strictness in following specific and concrete patterns. In fact, the structures detected in cluster ID 4 are depicted and analysed using a visualization tool (Gephi), and Figure \ref{fig:cl4} reports the predominant/common topology. This structure represents 13 of the 17 communities shared by the HDBSCAN and OPTICS in cluster ID 4. These communities are characterised by two transaction nodes connected by exactly two addresses, with a significant amount transferred directly from one input to one output without substantial splitting (the width of the edges in the graph). Furthermore, when labels are considered within this analysis, they show that the highest input in TX$\textsubscript{1}$ and the highest output in TX$\textsubscript{2}$ are both \textit{Exchange} addresses.

\section{Conclusion and Future Work}\label{sec:conclusion}
This work introduces an initial step toward the extraction of cryptocurrency mixers' \textit{modus operandi}. Specifically, it leverages address-transaction graphs to map mixer activities and applies graph-based analyses to identify topological similarities and patterns. This approach has been used to dissect \textit{Blender.io} activities, revealing that these activities follow specific and well-defined topological structures. Additionally, it highlights the pivotal role of \textit{Exchange} entities in mixer operations, raising concerns about their AML/KYC policies.

Although this work represents ongoing research, it provides an intriguing yet partial view of mixer activities. As a next step, we plan to include two additional dimensions to the methodology. The first focuses on improving the topological analysis by examining how mixer addresses interact within the same graphs, as they may form specific patterns to accumulate or split funds (e.g., aggregation networks, peeling chains, etc.). The second dimension aims to integrate economic aspects into the address-transaction graphs to determine whether mixers follow identifiable patterns in terms of received/sent amounts and temporal activity. By incorporating these elements, we seek to refine investigative approaches, provide more actionable intelligence.

\section*{Acknowledgment}
This work has been partially supported by the European Union's Horizon 2020 Research and Innovation Program under the project SAFEHORIZON (Grant Agreement No. 101168562).

\bibliographystyle{splncs04}
\bibliography{main}

\end{document}